\documentclass[11pt]{article}
\usepackage[latin1]{inputenc}
\usepackage[english]{babel}
\usepackage[namelimits]{amsmath}
\usepackage{amssymb}
\usepackage{amsmath}
\usepackage{amsthm}

\begin{document}
\title{Planckian corrections to the Friedmann flat equations from thermodynamics at the apparent horizon}
\author{
Stefano Viaggiu,\\
Dipartimento di Matematica,
Universit\`a di Roma ``Tor Vergata'',\\
Via della Ricerca Scientifica, 1, I-00133 Roma, Italy.\\
E-mail: {\tt viaggiu@axp.mat.uniroma2.it}}
\date{\today}\maketitle
\begin{abstract}
In this paper we use our recently generalized black hole entropy formula to propose a quantum version of the Friedmann 
equations. In particular, starting from the differential version of the first law of thermodynamics, we are able to find 
planckian (non commutative) 
corrections to the Friedmann flat equations. The so modified equations are formally similar to the ones present in Gauss-Bonnet gravity,
but in the ordinary 3+1 dimensions. 
As a consequence of these corrections, by considering negative fluctuations in the internal energy that are allowed by quantum field theory, our equations imply a maximum value both for the energy density $\rho$
and for the Hubble flow $H$, i.e. the big bang is
ruled out. Conversely, by considering positive quantum fluctuations, we found no maximum for $\rho$ and $H$. Nevertheless, by starting with an
early time energy density $\rho\sim 1/t^2$, we obtain a value for the scale factor $a(t)\sim e^{\sqrt{t}}$, implying a finite planckian universe 
at $t=0$, i.e. the point-like big bang singularity is substituted by a universe of planckian size at $t=0$. Finally, we found 
possible higher order planckian
terms to our equations together with the related corrections of our generalized Bekenstein-Hawking entropy.
\end{abstract}

\section{Introduction}

The Hawking's discovery \cite{1} concerning the radiation from a black hole represents an essential tool to study the relation from thermodynamics, 
black holes and quantum effects in curved spacetimes. In particolar, to a black hole in an asymptotically flat spacetime can be associated an
entropy \cite{2} $S_{BH}$ given by $S_{BH}=\frac{k_B A}{4 L_P^2}$, where
$k_B$ is the Boltzmann constant and $L_P$ the Planck length. This fundamental discovery opened the door to interesting investigations concerning
the thermodynamics of the universe as a whole both in general relativity and modified gravity 
(see for example \cite{7,3,4,4a,5,6,6a,8,9,10,11}) at the apparent horizon of the universe \cite{12,13}. 
Recently (see for example \cite{4,4a,5,6,7,8,9,10,11}), 
the interest was devoted to the derivation of Friedmann equations in general relativity and modified
gravity from the first law of thermodynamics at the apparent horizon. In all these derivations, thanks to the holographic principle
\cite{4,4a}, it is assumed for the entropy $S_h$ of the apparent horizon the
usual one of a static black hole $S_{h}=\frac{k_B A_h}{4 L_P^2}$, where
$A_h$ is the proper area of the apparent horizon. In a cosmological context apparent horizons are dynamical and non static. 
Moreover, in ordinary cosmology the situation is more involved since we have not at our disposal a way to calculate the dynamical degree
of freedom of the gravitational field.

Recently, \cite{14,15} we proposed a generalized expression for the entropy of black holes embedded in Friedmann universes.
This generilized formula can be applied, thanks to the holographic principle, to the apparent horizon 
and can be interpreted as the total entropy of our universe. In this formula a further term does apper proportional to 
the Hubble flow $H$ and describing the dynamical degrees of freedom of the gravitational field. As a consequence of this new proposal and
according to an old conjecture \cite{16},
the classical Friedmann flat universe has a time independent vanishing internal energy. When quantum effects are taken into account by a non 
vanishing Planck constant $\hbar$, a series expansion for the internal energy of a Friedmann flat spacetime can be performed.
 
In this paper, instead of write down the Friedmann equtions as a first law at the apparent horizon, starting from these new results
and from our proposed first law, we
propose a quantum (non commutative) ultraviolet modified version of the Friedmann equations.

In section 1 we review our generilezed Bekenstein-Hawking entropy formula for the entropy of the universe.
In section 2 we write down our first law for a Friedmann flat spacetime suitable to generilize the Friedmann equations and in section
3 we present our method applied to classical Friedmann spacetimes.
In section 4, we obtain an ultraviolet modification of the de Sitter spacetime and in section 
5 we generalize and motivate our method to any 
Friedmann flat spacetime together with a proposal for a non spatially flat spacetime. Section 6 is devoted to a study of 
possible higher order planckian corrections. Finally, section 7 is devoted to the conclusions and final considerations.

\section{Generalized Bekenstein-Hawking entropy at the apparent horizon}

The line element for a Friedmann universe is in comoving coordinates
\begin{equation}
ds^2=-c^2 dt^2+a^2(t)\left[\frac{{dr}^2}{1-kr^2}+r^2{d\Omega}^2\right],
\label{1}
\end{equation}
where $k=-1,0,+1$.
A Friedmann universe filled with matter-energy satisfying the weak energy condition is equipped with an apparent horizon at the proper 
areal radius $L_h$ given by
\begin{equation} 
L_h=\frac{c}{\sqrt{H^2+\frac{k c^2}{a(t)^2}}}.
\label{2}
\end{equation} 
It is customary to associate to the apparent horizon \cite{5} the Hawking temperature $T_h=c\hbar/(2\pi k_B L_h)$.
In particular, in the paper \cite{5} we have for the first time a derivation of the Friedmann equation from the first law at the apparent horizon:
\begin{equation}
T_{h}dS=dU_h+W_h dV_h,\;\;\;W_h=\frac{p-\rho c^2}{2},
\label{3}
\end{equation}
where $W_h$ is the work term and $U_h=c^4L_h/(2G)$ is the Misner-Sharp energy term (see also \cite{15}), $p$ the pressure and 
$\rho$ the energy density. 
The Friedmann equations are consistant with
(\ref{3}) provided that it is assumed for the apparent horizon the entropy $S_h=\frac{k_B A_h}{4 L_P^2}$. Generally, the entropy of the whole universe at the apparent horizon is assumed to be the sum of the entropy of the holographic screen ($S_h$) with the usual one of the matter inside
$L_h$. This approach is not absent from criticisms. For example, in \cite{17} it has been shown that there exist  
different heat sign conventions between the Clausius equation in \cite{5} (positive heat out sign convention) and the usual Gibbs equation
(positive heat in sign concention) for the matter inside $L_h$. This fact takes necessary a modified version of the first law for the matter content inside
$L_h$. Moreover, as pointed also in \cite{18} at the introduction, we have not at our disposal a way to calculate the dynamical degrees of
freedom for a non-static gravitational field. In particular, the assumption that $S_h$ 
is the entropy suitable also for non static black holes 
seems trather artificial since we expect a modification caused by the expansion of the universe, i.e. by the Hubble flow $H$.
The same criticism does apply to the usual setup for the entropy inside $L_h$.

In \cite{14,15}, by using suitable theorems for the formation of trapped surfaces for mass-energy concentration embedded in Friedmann
universes, we proposed a generalization for the entropy of a black hole of proper are $A_h$ and volume $V_h$ embedded in a Friedmann 
universe given by
\begin{equation}
S_{h}=\frac{k_B A_h}{4 L_P^2}+\frac{3k_B}{2c L_P^2}V_h H-
\frac{3k k_B}{4L_P^2}\frac{L_h V_h}{a(t)^2}.
\label{4}
\end{equation}
In the static case, expression (\ref{4}) reduces to the usual one. To calculate the expression for the entropy of the whole universe at
$L_h$ we use the holographic principle \cite{4,4a}. In fact, in the static case, the total entropy for a black hole is given by
$S_{ev}=\frac{k_B A_{ev}}{4 L_P^2}$, where $A_{ev}$ denotes the proper area of the event horizon, i.e. the entropy of the holographic
screen and there is not a contribution from the interior of the black hole. In a cosmological spacetime the event horizon becomes a 
teleological object that is not attainable for human observers. In this context the useful tool to identify black holes is provided by
apparent horizons (see \cite{12,13} and the discussion in \cite{14,15}). As a consequence, we can assume that the total entropy
(comprising the dynamical degrees of freedom) of the universe is given by the (\ref{4}) calculated at $L_h$ given by
(\ref{2}). In practice, we suppose that the entropy bound is exactly satured at the apparentr horizon \cite{14,15}.
In the following we mainly consider the flat case with $k=0$. Concerning the temperature, note that after writing the first law as
\begin{equation}
T_{h}dS_h=dU_h+W_h dV_h,
\label{7}
\end{equation}
the equation (\ref{7}) is left unchanged under a constant $\beta$ scaling of $T_h$
(i.e. $T_h\rightarrow\beta T_h, U_h\rightarrow\beta U_h, W_h\rightarrow\beta W_h$). We advantage of this to choose $\beta=1/2$. As a consequence of these setups we have, thanks to the (\ref{4}) with $k=0$:
\begin{eqnarray}
& & dU_h=\frac{c^4}{2G}dL_h+\frac{c^3}{2G}L_h^2dH,\label{8}\\
& & W_h=\frac{3 c^3 H}{8\pi G L_h},\;T_{h}=\frac{\hbar c}{4\pi k_B L_h}.\label{9}
\end{eqnarray}
In this way, with $\beta=1/2$ the first term in (\ref{8}) is nothing else but the Misner-Sharp mass and $W_h=\rho$ 
after using the Friedmann equation $H^2=(8\pi G\rho)/3$. Also note that after taking $L_h=c/H$ we have 
$U_h=const$\footnote{The Misner-Sharp mass in (\ref{8}) is exactly balanced by the added term describing the dynamic degrees of freedom
	of the gravitational field.}. 
As discussed in \cite{15}, this implies $U_h=0$ in the flat case, i.e. the Friedmann flat universes, according to 
\cite{16}, all have zero internal energy. The fact that only in the spatially flat case we obtain $U_h=0$ strongly motivates our 
proposal (\ref{4}) as the entropy of the universe at its 'thermodynamic' radius $L_h$.
Equations (\ref{7})-(\ref{9}) are the starting point for the investigations of this paper.

\section{Friedmann flat equations from the first law of thermodynamics}

For a Friedmann flat spacetime the differential first law can be written as
\begin{equation}
T_{h}dS_h=dU_h+c^2\rho\;dV_h,
\label{10}
\end{equation}
in such a way that only thermodynamic quantities appear in the first law. We assume that the differential 
form (\ref{10}) of the first law remains left unchanged also when quantum corrections are taken into account, that is in line with the standard 
assumptions present in the literature (see for example \cite{5,6,9,10,11}). 
We advantage of the presence of
$\rho$ in (\ref{10}) to obtain the field equations by the knowledge of $S_h$ and $U_h$. To start with, we introduce 
the free Helmholtz energy $F_h=U_h-T_h S_h$. At the apparent horizon, the proper volume $V_h$ and the temperature
$T_h$ are not independent quantities: they satisfy the equality
\begin{equation}
T_{h}^3 V_h = \frac{1}{48{\pi}^2}{\left(\frac{c\hbar}{k_B}\right)}^3.
\label{11}
\end{equation}
Note that, as pointed in \cite{17}, $T_h$ is the temperature of the holographic screen that is generally different from the one
$T_u$ of the universe inside $L_h$. In practice $T_h^3 V_h=const$, as happens for an isoentropic radiation field.
As a consequence, we obtain
\begin{equation}
dF_h=-\rho V_{h,T}(T)dT-S_h dT,\;\;V_{h,T}(T)=-{\left(\frac{c\hbar}{k_B}\right)}^3\frac{1}{16{\pi}^2 T_h^4},
\label{12}
\end{equation} 
where "$,$" indicates partial derivative.
From (\ref{12}) we can obtain $\rho$ in terms of the total derivative of $F_h$ with respect to $T_h$, i.e. 
$\frac{dF_h}{dT_h}$ and $S_h$. We have
\begin{equation}
c^2\rho=-\left(\frac{dF_h}{dT_h}+S_h\right)\frac{1}{V_{h,T}}.
\label{13}
\end{equation}
We can check that, by taking in (\ref{13}) the expression for $S_h$ dictated by a Friedmann flat universe 
(equation (\ref{4}) with $k=0$) 
\begin{equation}
S_h=\frac{3c^5\hbar}{16\pi G k_B}\frac{1}{T_h^2},\;\;T_h=\frac{\hbar H}{4\pi k_B},\;\;F_h=-T_h S_h,
\label{14}
\end{equation}
we obtain the classical Friedmann equation 
\begin{equation}
\rho=\frac{6{\pi} k_B^2 T_h^2}{G{\hbar}^2}.
\label{15}
\end{equation}
In the next three sections we use this simple technique to generate planckian corrections to the Friedmann equations.

\section{Planckian corrections for a cosmological de Sitter universe}

With only the constants $c,G,\Lambda$ (where $\Lambda$ is the cosmological constant) in the Einstein equations we get \cite{15}
$U_h=const=0$ for Friedmann flat spacetimes. Conversely, with the introduction of $\hbar$ we can have
\begin{equation}
U_h\sim \frac{G^{n-1}{\hbar}^n}{c^{3n-4}}{\Lambda}^{\left(n-\frac{1}{2}\right)},\;\;\;n\in N.
\label{16}
\end{equation}
We can thus build the following series expansion for $U_h$:
\begin{eqnarray}
U_h&=&c\hbar\sqrt{\Lambda}\left[\overline{c_0}+\overline{c_2} L_P^2\Lambda+\overline{c_4}L_P^4{\Lambda}^2+\cdots+\overline{c_{2n}}L_P^{2n}{\Lambda}^n+\cdots\right]\nonumber\\
&=&c\hbar\sqrt{\Lambda}\sum_{n=0}^{\infty}\overline{c_{2n}}L_P^{2n}{\Lambda}^n,\;\;\overline{c_{2n}}\in R.
\label{17}
\end{eqnarray}
Note that, apart from the unknown adimensional constants $\overline{c_{2n}}$, the higher order planckian corrections depend on the product
$L_P^{2n}{\Lambda}^n$ and thus are relevant for $\Lambda\sim 1/L_P^2$.
In the following we consider only the first order correction where 
$U_h=\overline{c_0} c\hbar\sqrt{\Lambda}$. Section 6 is devoted to the study of higher order 
planckian corrections. Moreover, to the non vanishing variation $U_h\sim\sqrt{\Lambda}$ can be associated, thanks to (\ref{11}), 
a constant entropy (radiation like field) given by
\begin{equation} 
S_0=w_0\frac{16\sigma}{3c}V_h T_h^3,\;\;\;\sigma=\frac{{\pi}^2 k_B^4}{60{\hbar}^3 c^2},
\label{18}
\end{equation}
where $w_0=1$ for pure radiation.
We have:
\begin{equation}
F_h=4\pi\sqrt{3}\overline{c_0} k_B T_h-S_0 T_h-\frac{3c^5\hbar}{16\pi G k_B T_h},\;\;T_h=\frac{c\hbar}{4\pi k_B}\sqrt{\frac{\Lambda}{3}}.
\label{19}
\end{equation}
From (\ref{13}) we obtain
\begin{equation}
c^2{\rho}_{\Lambda}=
\frac{{(4\pi)}^3\overline{c_0}k_B^4\sqrt{3}}{c^3{\hbar}^3}T_h^4+
\frac{3c^2{(4\pi)}^2 k_B^2}{8\pi G{\hbar}^2}T_h^2.
\label{21}
\end{equation}
Note that, thanks to the modified expression for $F_h$, a further term proportional to $T_h^4$ (radiation-like term) does appear in
(\ref{21}). A naive interpretation of (\ref{21}) suggests a Friedmann equation $H^2=8\pi G/3{\rho}_{\Lambda}$ with 
${\rho}_{\Lambda}={\Lambda}_{eff}c^2/8\pi G$ and ${\Lambda}_{eff}=\Lambda+2\overline{c_0}\sqrt{3}/9{\Lambda}^2 L_P^2$. 
Also in modified gravity we expect that (see \cite{6,9,10}) $T_h\sim H$, i.e. $T_h\sim\sqrt{\Lambda}$. But with the equation
${\rho}_{\Lambda}={\Lambda}_{eff}c^2/8\pi G$ we have obviously $T_h\sim\sqrt{{\Lambda}_{eff}}$, with 
${\Lambda}_{eff}\neq\Lambda$, a contradiction. With this interpretation, backreaction is not taken into account.
A self consistent way to read equation (\ref{21}) is to see this as a modified Friedmann equation. We start with a 'bare' cosmological
constant $\Lambda$. Thanks to quantum effects depicted by the first term in (\ref{17}) we have a modified 
observed cosmological constant
${\Lambda}_{obs}=\overline{\Lambda}$ with $T_h=\hbar\overline{H}/(4\pi k_B)$:
\begin{equation}
\frac{8\pi G{\rho}_{\overline{\Lambda}}}{3}={\overline{H}}^2+
\frac{2\overline{c_0}L_P^2}{c^2\sqrt{3}}{\overline{H}}^4.
\label{22}
\end{equation} 
In practice, thanks to quantum effects we have $\Lambda\rightarrow\overline{\Lambda}$. As customary in the ordinary renormalization group approach
(see \cite{19}), the relation between $\Lambda$ and $\overline{\Lambda}$ is unobservable, but in the modified Friedmann equation (\ref{22})
${\rho}_{\overline{\Lambda}}$ is the measured energy density. The correction term becomes important for a huge $\sim 1/L_P^2$ value of
the observed Hubble flow $\overline{H}$. Our approach is not equivalent to the introduction of an ad hoc holographic running
cosmological constant $\Lambda=\Lambda(H)$ coupled with the Hubble flow $H$, where the usual Friedmann equation
$H^2=8\pi G/3{\rho}$ is assumed together with an unusual coupling between matter and geometry
\footnote{Although the equation $H^2=\rho(H)$ is allowed, it represents a conceptual violation 
of the general relativity where $\rho$ is assigned and its behaviour determines the geometry, namely $H$. In this regard, the introduction of
a density $\rho(H)$ can be better associated to a modification of the Einstein tensor $G_{\mu\nu}$ rather than the energy momentum
tensor $T_{\mu\nu}$.}.   
In the next section we write down the equation (\ref{22}) to generic Friedmann flat spacetimes.

\section{Building planckian corrections to Friedmann flat equations}

As stated above, for any Friedmann flat spacetime we have $U_h=0$. For a de Sitter spacetime, also by introducing $\hbar$ with 
the series (\ref{17}), we have $U_h=const\neq 0$. The situation changes for a generic Friedmann flat spacetime. In fact, we can
always suppose, in presence of a non vanishing cosmological constant that $U_h\sim\sqrt{\Lambda}$ but now 
$\Lambda$ is not proportional to $T_h$. Fortunately, quantum mechanics comes in help. A certain energy displacement
$\Delta E$ is allowed provided that during a time $\Delta t$ we have $\Delta t\Delta E\geq \hbar/2$. By considering again the first term in 
the expansion (\ref{17}) we have:
\begin{equation}
4\pi k_B|c_0|\sqrt{3}\Delta t\Delta T_h \geq \frac{\hbar}{2},               
\label{23}                
\end{equation}
where $|c_0|$ to allows possible negative fluctuations of a quantum field (see for example \cite{20}).
Thanks to the (\ref{14}) we have
\begin{equation}
|c_0|\sqrt{3}\Delta t\Delta H\geq \frac{1}{2}.
\label{24}
\end{equation}
Since we find planckian corrections, it is natural to suppose a quantum spacetime \cite{21,22,23,24} 
at the Planck length $L_P$. In this context, we can consider physically motivated spatetime uncertainty relations (STUR)
\footnote{See \cite{21} for the STUR in minkowski spacetime in the Newtonian approximation, \cite{25} for the STUR in minkowski spacetime without
Newtonian approximation and \cite{26} for the generalization to Friedmann flat spacetimes.} between time and spatial coordinates.
In the Friedmann flat case \cite{26}, the chosen coordinates are the cosmic time $t$ and the spatial coordinates 
${\eta}^i=a(t) x^i$, where $x^i$ are the usual cartesian coordinates. Since we consider the universe at its thermodynamic radius $L_h$, it is 
natural to consider only localizing states ${\omega}_s$ that have spherical symmetry, in particular at very early times near the
big bang where the universe was effectively planckian. In these states  ${\omega}_s$
the uncertainty of the coordinates all have the same 
magnitude, i.e.  $c\Delta_s t\sim \Delta_s{\eta}^1=\Delta_s{\eta}^2=\Delta_s{\eta}^3=\Delta_s\eta$. 
Under these reasonable assumptions the STUR in \cite{26} imply a lower bound for
$\Delta_s t$ and $\Delta_s\eta$, namely $c\Delta_s t\sim\Delta_s\eta\sim L_P$. By applying this lower bound in (\ref{24}) we obtain 
an upper bound for $\Delta H$ given by
\begin{equation}
\Delta_s H_{max}=\frac{c}{2|c_0|\sqrt{3}L_P}.
\label{25}
\end{equation}
Hence, quantum fluctuations are available thanks to the Heisenberg uncertainty relation between time and energy, and coupled to planckian effects 
we expect an upper limit for the uncertainty of $H$ (and $T_h$). In fact, the STUR in \cite{26}, in presence of an infrared 
cutoff\footnote{This infrared cutoff is the particle horizon $L_{par}$ in \cite{26} and the apparent horizon $L_h$ in this paper 
with $L_{par}\sim L_h$} imply an upper bound for the mean value of 
$\omega_s(H)$ given by $\omega_s(H)=\frac{c}{\sqrt{3}L_P}$, in agreement with 
the (\ref{25}) with $|c_0|\sim 1/2$. Also note that by taking as infrared limit for the inequality (\ref{24}) the apparent horizon of our universe, we obtain a lower limit for the Hubble flow that is, at present day and with $|c_0|$ of the order of unity, of the same order of the one effectively
measured. This is a known fact in the holographic dark energy context, but in our context can have a new interpretation.

As a consequence of these reasonings, we can use again the expansion (\ref{17}) for $U_h$ expressed in terms of $T_h$, with the first term 
given by $U_h=4\pi\sqrt{3}c_0 k_B T_h$. By considering an observed energy density (including the 
cosmological constant ) depicted by $\sum_i{\overline{\rho}_i}={\overline{\rho}}$ and
using exactly the same technique applied for (\ref{22}) we obtain
\begin{equation}
\frac{8\pi G}{3}{\overline{\rho}}={\overline{H}}^2+
\frac{2\overline{c_0}L_P^2}{c^2\sqrt{3}}{\overline{H}}^4.
\label{26}
\end{equation} 
The term $\sim{\overline{H}}^4$ in (\ref{26}) is in line with the ultraviolet modifications of the general relativity present 
in \cite{30} by using
the Arnowitt-Deser-Misner formalism and also in brane word scenario (see for example \cite{31}). However, in our approach the modifications
of the Friedmann equations arise from quantum effects driven by the thermodynamics at the apparent horizon, 
while in \cite{30} are a geometrically based consequence 
of the modification from the onset of the Hilbert-Einstein action. In our approach naturally arises the Planck length
indicating a non commutative nature of the ultraviolet modification of the Friedmann equations. 
Moreover, with our approach we can obtain a physically sound interpretation of the geometric based qpproaches, as the one in \cite{30},
in terms of a variation of the internal energy $U_h$ and the entropy $S_h$ of the universe thanks to quantum fluctuations.

Note that our modified Friedmann equation has the same form of the one present in Gauss-Bonnet gravity. In the Gauss-Bonnet
gravity the constant multiplying the term $H^4$ is vanishing in ordinary $3+1$ dimensions, while our modification is an ultraviolet
correction acting on a four dimensional manifold. This formal analogy implies for the other Friedmann flat equation the 
following expression: 
\begin{equation}
c^2\left(1+\frac{4\overline{c_0}L_P^2}{c^2\sqrt{3}}H^2\right)H_{,t}=-4\pi G(c^2{\overline{\rho}}+
{\overline{p}}),
\label{27}
\end{equation}
where ${\overline{p}}$ is the pressure.

Also in \cite{27} an expression similar to (\ref{26}) was obtained in an attempt to obtain the  modified Friedmann equations in
Horava-Lifshitz gravity from the first law of thermodynamics at the apparent horizon. The author of \cite{27} obtains 
this equation that is different from the one present in Horava-Lifshitz modified gravity, deducing that it is not possible to obtain the correct
Friedmann expression in Horava-Lifshitz gravity from the first law. We stress again that our derivation of 
(\ref{26}) has been obtained from our generized Bekenstein-Hawking formula without forcing the first law at $L_h$ to be exactly the
Friedmann equations with the rather questionable assumption $S_h\sim A_h/4$. 

In a quantum context, equations (\ref{26}) and (\ref{27}) can be seen as effective field equations where the quantities 
${\overline{\rho}},{\overline{p}},{\overline{H}}$ are mean values calculated on some quantum state $s$, i.e.
${\overline{\rho}}=\omega_s(\rho)=\omega_s(T_{00}), {\overline{H}}=\omega_s(H)$. 

Obviously, a modification of the Friedmann equations implies a modification of our generized entropy formula
(\ref{4}) $S_{h}=\frac{k_B A_h}{4 L_P^2}+\frac{3k_B}{2c L_P^2}V_h H$. After a simple algebra it is easy to see that the planckian 
modified expression ${\overline{S}}_h$ is
\begin{equation}
{\overline{S}}_h=\frac{k_B A_h}{4 L_P^2}+\frac{3k_B}{2c L_P^2}V_h{\overline{H}}
\left[1+\frac{w L_P^2}{1080\pi c^2}{\overline{H}}^2\right].
\label{28}
\end{equation}
In section 6 we briefly discuss the role of the other terms in the series expansion (\ref{17}).

Following the analogy with the Gauss-Bonnet Friedmann equations, we can propose a modification of the non spatially flat 
Friedmann equations. This could be easily done performing in (\ref{26}) and (\ref{27}) the transformation
${\overline{H}}^2\rightarrow {\overline{H}}^2+\frac{k c^2}{a(t)^2}$. However, at present we have not further arguments to justify
this generalization and we leave to future investigations this issue.

From equation (\ref{26}) we can solve with respect to ${\overline{H}}^2$ to obtain:
\begin{equation}
{\overline{H}}^2=\frac{c^2\sqrt{3}}{4{\overline{c_0}} L_P^2}
\left[-1+\sqrt{1+\frac{64\pi{\overline{c_0}}}{3\sqrt{3}}\frac{\overline{\rho}}{{\rho}_P}}\right],
\label{29}
\end{equation}
where ${\rho}_P$ is the Planck density. Obviously, in the limit $L_P\rightarrow 0$ we recover the classical Friedmann equation.

Note that equation (\ref{29}) is defined for both ${\overline{c_0}}>0$ and ${\overline{c_0}}<0$.
In the next two subsections we study our equations (\ref{26}) and (\ref{27}) in the cases ${\overline{c_0}}>0$ and
${\overline{c_0}}<0$.

\subsection{Positive quantum fluctuations}

In this subsection we consider the case with ${\overline{c_0}}>0$. First of all, we can obtain the behaviour of
(\ref{29}) for low densities $\overline{\rho}<<(3\sqrt{3}{\rho}_P)/(64\pi{\overline{c_0}})$. We have
\begin{equation}
{\overline{H}}^2=\frac{8\pi G}{3}\overline{\rho}
	\left[1-\frac{\sqrt{3}\pi 2^7{\overline{c_0}}}{9}\frac{\overline{\rho}}{{\rho}_P}\right].
\label{30}
\end{equation}
Equation (\ref{30}) looks like the well known Friedmann flat equation present in loop quantum bouncing cosmologies
(see \cite{28} and reference therein). Despite this nice analogy, the equation (\ref{29}) does not admit a turning point
where ${\overline{H}}=0$. Moreover, from (\ref{29}) it is evident that for ${\overline{c_0}}>0$
we have not an upper bound for the density and the Hubble flow. Nevertheless, we expect that some planckian effect can arise near the 
classical big bang singularity. To this purpose, consider the usual setup near the big bang for ${\overline{\rho}}$:
${\overline{\rho}}=3\alpha/(8\pi G t^2)$ for some positive constant $\alpha$ of the order of unity. For 
$t\rightarrow 0$, the dominant term of (\ref{30}) it gives:
\begin{equation}
\overline{H}=\frac{\sqrt{B}}{\sqrt{t}},\;
B=\frac{2\sqrt{3}c^2}{L_P^2\sqrt{\overline{c_0}}}
\sqrt{\frac{\alpha}{8\pi G\sqrt{3}{\rho}_P}},\;\;a(t)=a_0e^{2\sqrt{B t}}.
\label{31}
\end{equation}
Although the energy density follows the classical behaviour for $t\rightarrow 0$, we have that $a(t=0)=a_0\neq 0$.
This means that, thanks to quantum effects at planckian level, the point like universe at $t=0$ is substituted by a 
spherical planckian but finite ball. This is in agreement with the fact that the STUR in a Friedmann flat spacetime imply, for a 
spherical localizing state \cite{26} that $\Delta_s\eta\geq L_P$, i.e. a fuzzy sphere. A modification of the nature of the
big bang singularity caused by non commutative effects can also be found in \cite{29}.

\subsection{Negative quantum corrections: big bang is ruled out}

In the case ${\overline{c_0}}<0$ we expect a stronger manifestation of quantum planckian effects with respect to the case
with ${\overline{c_0}}>0$. To start with, note that  the existence condition for the root in (\ref{30}) implies a maximum value
for $\overline{\rho}$ given by:
\begin{equation}
{\overline{\rho}}_{max}=\frac{3\sqrt{3}{\rho}_P}{64\pi|\overline{c_0}|}.
\label{32}
\end{equation} 
The maximum value can be made the one obtained in loop quantum gravity \cite{28}, namely 
${\overline{\rho}}_{max}\simeq 0.41{\rho}_P$ with $|\overline{c_0}|\simeq 0.06$. Obviously, thanks to (\ref{30}),
for ${\overline{\rho}}_{max}$ given by (\ref{32}) we have an ${\overline{H}}_{max}$ given by
\begin{equation}
{\overline{H}}_{max}=\frac{c\;3^{\frac{1}{4}}}{2\sqrt{|\overline{c_0}|}\;L_P}.
\label{33}
\end{equation}
Hence, as in \cite{30},
our modified equation (\ref{29}) with ${\overline{c_0}}<0$ is free from the initial singularity at $t=0$ and big bang is ruled out.
Moreover, equation (\ref{27}) implies for ${\overline{H}}_{max}$ a de Sitter solution with $c^2\rho=-p$. 
The  upper bound in (\ref{33}) is in agreement with the one present in \cite{26} with $\overline{c_0}$ of the order of unity.
With $|\overline{c_0}|=\sqrt{3}/4$ we have a universe at $t=0$ given by a ball of proper radius $L_P$.
This interesting result means that 
non commutative effects with negative fluctuations generates an effective positive cosmological constant that takes the place of the 
big bang paradigm. Hence, a Friedmann flat universe can be emerged from a minkowski spacetime by quantum 
fluctuations. The fact that, thanks to our new proposal \cite{14,15}, both the 'unerturbed' (classical) 
minkowski and Friedmann flat spacetimes have zero internal energy, takes possible this interesting scenario. 

As a final remark of this section, consider a universe filled with dark matter ${\overline{\rho}}_m$ after recombination. We can approximate the
(\ref{26}) with ${\overline{c_0}}<0$:
\begin{equation}
{\overline{H}}^2=\frac{8\pi G}{3}{\overline{\rho}}_m
\left[1+\frac{\sqrt{3}\pi 2^7|{\overline{c_0}}|}{9}\frac{{\overline{\rho}}_m}{{\rho}_P}\right].
\label{34}
\end{equation}
A similar expression can be found in \cite{32}.
After dividing the (\ref{34}) for ${\overline{H}}^2$ and introducing the usual density parameter ${\overline{\Omega}}_m$
we get
\begin{equation}
1={\overline{\Omega}}_m+{\overline{\Omega}}_m
\frac{{\overline{\rho}}_m}{{\rho}_P}|{\overline{c_0}}|\pi 2^7\frac{\sqrt{3}}{9}.
\label{35}
\end{equation}
Formula (\ref{35}) shows that, also away from the primordial planckian era $t\simeq t_p$ and on cosmological scales (apparent horizon),
non commutative effects generate an effective cosmological constant $\tilde{\Lambda}$ given by
\begin{equation}
\tilde{\Lambda} = \frac{8\pi G}{c^2}|{\overline{c_0}}|\frac{\sqrt{3}}{9}\pi\;2^7\frac{{\overline{\rho}}_m^2}{{\rho}_P}.
\label{36}
\end{equation}
Unfortunately, with $|{\overline{c_0}}|$ of the order of unity, $\tilde{\Lambda}$ is too small to explain the actual value of the cosmological
constant. To this purpose a very huge unnatural value for $|{\overline{c_0}}|$ is necessary. Nevertheless, it is physically interesting that also at 
infrared cosmological scales planckian quantum effects can exist. Only near the planckian era the non commutative effects are so strong that a
huge cosmological constant comes in action with the born of our dynamical universe. 

\section{Higher order corrections}

In this section we study some consequences of possible higher order corrections in the expansion $(\ref{17})$. In terms of the temperature
$T_h$ we have
\begin{equation}
U_h=4\pi k_B\sqrt{3} \overline{c_0} T_h+\frac{\overline{c_2} L_P^2}{c^2 {\hbar}^2}{\left(4\pi k_B\sqrt{3}\right)}^3 T_h^3+\cdots
\frac{\overline{c_{2n}} L_P^{2n}}{c^{2n} {\hbar}^{2n}}{\left(4\pi k_B\sqrt{3}\right)}^{2n+1} T_h^{2n+1}.
\label{37}
\end{equation}
With the same technique of section 4 we obtain
\begin{equation}
\frac{8\pi G}{3}{\overline{\rho}}={\overline{H}}^2+
\frac{2\overline{c_0}}{c^2\sqrt{3}}L_P^2{\overline{H}}^4+
\frac{c_2 6\sqrt{3}}{c^4}L_P^4{\overline{H}}^6+higher\;\;orders,
\label{38}
\end{equation} 
where $c_2$ is defined in terms of $\overline{c_2}$ and with the entropy term $S_0 T_h$ in
(\ref{19}) substitued with $S_{2} T_h^3$ with $S_2=w_2\frac{16\sigma}{3c}V_h T_h^3$ where
\begin{equation}
3\overline{c_2}{(4\pi\sqrt{3})}^3-\frac{w_2}{270}=3c_2{(4\pi\sqrt{3})}^3.
\label{39}
\end{equation}
and so on.
We obtain an expansion where only even powers of $H$ appear, formally preserving  general covariance 
(see \cite{19} and references therein). 

In the case where $\{\overline{c_0},c_2\}<0$, similar conclusions outlined in subsection 5.2 are valid. Consider now the case of a generic
correction up to ${{c_{2n}}}L_P^{2n+2}{\overline{H}}^{2n+4}$ with ${c_{2n}}>0$. By considering again 
${\overline{\rho}}=3\alpha/(8\pi G t^2)$ we obtain that the term $\sim {\overline{H}}^{2n+4}$ will be dominant in the limit
$t\rightarrow 0$. From (\ref{38}) we obtain $\overline{H}\sim\frac{1}{t^{\frac{2}{2n+4}}}$ in this limit.\\ 
Under the hypothesis that there exists an
$n^*\in N$ such that $\forall n>n^*, {c_{2n}}>0$
and performing the limit $n\rightarrow\infty$ , we have $H\rightarrow constant$, i.e. a de Sitter universe and once again big bang is
ruled out, but this time we must sum all contributions in the series expansion (\ref{38}). In this case the conclusions of subsection 5.2 
does not apply and a small cosmological constant could arise compatible with its small observed value. This can be matter for further 
investigations.

\section{Conclusions and final remarks}

In this paper we presented a simple technique to obtain planckian modifications to the Friedmann equations by 
means of the first law at the apparent horizon
we recently presented \cite{14,15}. We mainly considered the spatially flat case. In this regard, the presence of 
$\rho$ in (\ref{10}) permit us to obtain the energy density as a function of the free Helmholtz energy $F_h$ and $S_h$, namely equation
(\ref{13}). In the Friedmann flat case \cite{26}, our generalized entropy implies $U_h=0$. However, when quantum fluctuations are considered, 
a non trivial expression for $U_h$ can be obtained. This non trivial expression for $U_h$ implies a modifications of the Friedmann equations and in turn a generalization of our entropy formula (\ref{4}). The expression of our modified Friedmann flat equations is very similar to the one obtained
in Gauss-Bonnet or Lovelock gravity, but our ultraviolet corrections act on ordinary $3+1$ dimensions. 
Obviously, the modification we obtained of Friedmann flat equations appeared in different contexts in the literature
(see for example \cite{30,31,32} and references therein), but in our context naturally does appear the Planck length as a measure of the 
deviation from the classical Friedmann equations, by suggesting a non commutative nature of the ultraviolet corrections. Moreover, 
the study of section 6 indicates that the ultimate form of the Friedmann equations could be done in terms of a series expansion 
of even power of $\overline{H}$. 

It is also interesting that our Friedmann
equation (\ref{29}) allows to consider both positive and negative fluctuations in the internal energy $U_h$. For positive fluctuations and low densities with respect to the Planck one ${\rho}_P$, 
the equation (\ref{29}) reduces to a form very similar to the one found in loop quantum gravity \cite{28}, but the non approximated expression
(\ref{29}) does not allow a maximum for ${\overline{\rho}}$
and ${\overline{H}}$. However, by considering positive fluctuations with a singular ${\overline{\rho}}$ given by the usual expression
${\overline{\rho}}\sim 1/t^2$, we found that $a(t)\sim e^{\sqrt{t}}$ for $t\rightarrow 0$. 
This fact means that also by starting with a singular energy density and 
Hubble flow, the universe at $t=0$ has finite Planckian dimensions,  and the point-like universe at $t=0$ is substituted by a finite
planckian ball. By considering negative fluctuations, our effective equations imply a maximum allowed value for both
${\overline{\rho}}$ and ${\overline{H}}$ that are in agreement with the ones obtained in \cite{26}. In this case big bang is ruled out by 
non commutative effects that are expected to be dominant for $t\rightarrow 0$. It is also interesting that our equations
(\ref{26}) and (\ref{27}) with ${\overline{c_0}}<0$ admits a de Sitter phase with ${\overline{H}}=H_{max}$ when
${\overline{\rho}}={\rho}_{max}$. This is not in agreement vith a 
bouncing non singular cosmology present in loop quantum gravity \cite{28} where ${\rho}_{max}$ corresponds to $H=0$. Conversely, this
is in agreement with
a universe born from a quantum minkowski spacetime.
In fact both the unperturbed minkowski and Friedmann flat spacetimes have zero internal energy $U_h$. Quantum fluctuations depicted in section
4 can thus generate a dynamical Friedmann flat universe with $H_{max}$ and can explain why we live in a Friedmann flat universe, while the 
bouncing scenario does not solve this fundamental cosmological issue. This fact is certainly matter for future investigations. Another line of research is to solve equations (\ref{26}) and (\ref{27}) to search some cosmological signature or relic provided by non commutative effects.

\section*{Acknowledgements} 
I would like to thank Luca Tomassini for interesting and useful discussions concerning the quantum nature of the spacetime.

\end{document}